\documentclass[aip,reprint]{revtex4-1}
\usepackage{graphicx}
\usepackage{listings}
\usepackage{picins}
\usepackage{latexsym}
\draft % marks overfull lines with a black rule on the right

\begin{document}

%\preprint{}

\title{Crystallization and non-crystallization of Lennard-Jones particles studied by molecular dynamics simulation} %Title of paper

% repeat the \author .. \affiliation  e$T_{c}$. as needed


\author{Hui Zhang}
\email[Electronic mail:]{zhope@scut.edu.cn}
\noaffiliation
%\collaboration{Zhongwu Liu, Xichun Zhong, Dongling Jiao, Wanqi Qiu}
\author{Zhongwu Liu}
\noaffiliation
\author{Xichun Zhong}
\noaffiliation
\author{Dongling Jiao}
\noaffiliation
\author{Wanqi Qiu}
\noaffiliation
%\thanks{}
\affiliation{School of Materials Science and Engineering, South China University of Technology, Guangzhou 510640, People's  Republic of China}

% Collaboration name, if desired (requires use of superscriptaddress option in \documen$T_{c}$lass). 

\date{\today}

\begin{abstract}
What lattice Lennard-Jones (LJ) solid favors, the lattice identification of simulated system and the microstructures of liquid and non-crystalline solid are three important questions in condensed physics and material science and are addressed in this paper. Both the crystallization and non-crystallization of LJ particles have been investigated by molecular dynamic (MD) simulation without setting any initial Bravais lattice. To identify the Bravais lattice of simulated system, two distribution functions of both the angles between one particle and its nearest neighbors and the distances between particles have been proposed. The final identification can be made by comparing these two calculated distribution functions with those of ideal Bravais lattices and checking the particle arrangement of simulated system. Our results have shown that simulated systems show either the face-centered cubic (fcc) lattice or the hexagonal close-packed (hcp) lattice. The microstructure of non-crystalline system is similar to that of LJ liquid at a temperature near the crystallization temperature, and shows no order of the second nearest neighbors in comparison with that of crystalline system. This paper has proposed a new way of investigating the microstructure of material and its evolution, and paved the way for MD simulation of large scale particle system consisting of more than one million particles.
\end{abstract}

%\pacs{77.80.Jk; 77.65.Bn; 77.80.Dj; 77.80.Fm}% insert suggested PACS numbers in braces on next line

\maketitle %\maketitle must follow title, authors, abstract and \pacs

\section{Introduction}
The classic Lennard-Jones (LJ) potential consists of a repulsive term and an attractive term and is commonly used to describe the interactions of rare gases \cite{Lennard-1}. Later it was used for crystalline solids and theoretical calculations indicated that the hexagonal close-packed (hcp) lattice has the lowest energy and its energy is 0.01\% lower than that of the face centered cubic (fcc) lattice \cite{Kihara-2,Barron-3}. The small difference in their energies may have an effect on their existence when cooled from the liquid. With the development of computer technology, LJ potential has been largely used in computer simulations as an interatomic potential. The simulated results of LJ solid have shown that the liquid-crystalline phase transition occurred but the lattice of simulated system is difficult to identify \cite{Mandell-4}. LJ solid may show the fcc lattice due to lattice defect although the hcp lattice is energetically more favorable \cite{Waal-5}. What lattice LJ solid favors has no clear answer so far and no simulations have provided solid evidences for it. However, this is an important question because LJ potential can be a starting point for investigating and constructing the interatomic potential for crystalline solids if we can identify the lattice of LJ solid. This can enrich our knowledge about the structure of material and the interaction between particles. Molecular dynamics (MD) simulation is the best way for our investigation and this leads to our second important question. For MD and other computer simulations, there is no effective way to identify the lattice of simulated system. During the post-treatment of simulated results, we can calculate the radial distribution function $g(r)$, the number of the nearest neighbors, and so on. But these calculated results are not helpful for the lattice identification of simulated system and sometimes are misleading. For instance, the number of the nearest neighbors is often larger than 12 and is incorrect. The lattice identification of simulated system is so important that the simulation cannot go further if we don’t know the lattice of simulated system. Third, we know little about the microstructures of liquid and non-crystalline system although great efforts have been made \cite{Wang-6}. However, if LJ crystalline solid can be formed by cooling LJ particles from the liquid, then both the liquid-crystalline and liquid-non-crystalline phase transitions can be reproduced by MD simulations. Thus by analyzing the simulated results, we can learn a lot about the microstructure of the liquid, the evolution of atomic arrangement of simulated system at the crystallization temperature, and the microstructure of non-crystalline solid. The information is helpful to understand the crystallization and non-crystallization.\\
\indent In this paper, the above three questions are solved by investigating both the crystallization and non-crystallization of LJ particles. MD simulations were carried out without setting any initial Bravais lattice and we proposed a new method for lattice identification of simulated system. First, we obtained LJ crystalline solid and made the lattice identification. Second, we investigated the microstructures of simulated system at the liquid state and crystalline state, and the sudden change of the atomic arrangements accompanying with the liquid-crystalline phase transition. Third, we obtained the non-crystalline system by adjusting the cooling rate and investigated the microstructure of simulated system and made a comparison with that of crystalline system.\\
\section{Simulation and method}
We introduce the classic LJ potential to describe the interatomic coupling, and LJ potential can be written as 
\begin{eqnarray}
U(r)=4\epsilon \left(\left(\frac{\sigma}{r}\right)^{12}-\left(\frac{\sigma}{r}\right)^{6}\right)
\end{eqnarray}
where $\epsilon$ is the depth of the potential well, $\sigma$ is the finite distance at which the inter-particle potential is zero, and $r$ is the distance between particles. The simulation was carried out with the aid of LAMMPS \cite{Plimpton-7}. In the simulation,  $\sigma$=1.0, and the distance of cutoff $r_{c}$=2.5. The LJ units were used, and the periodic boundary conditions were applied. The number of particles was 1000, and the mass of the particle was 100. We did not set any initial Bravais lattice. The particles were created randomly in the simulation box and then an energy minimization procedure followed. For every value of $\epsilon$, the initial temperature $T_{0}$=0.9$\epsilon$, and $\epsilon$=10-2000. We set the timestep as 0.001. At  $T_{0}$, NPT dynamics was implemented for $10^{6}$ timesteps, and then the temperature was decreased by  $T_{0}/n$, for $n$=40, 50, and 60. At every following temperatures $T$, NPT was carried out for a time of $t$ timesteps, and $t$=1$\times$10$^{3}$-1$\times$10$^{7}$. The pressure was always zero in the simulation.  Details can be found in in-script in Appendix. The visualization of simulated results was done with the aid of VESTA \cite{Momma-8}. The equilibrium nearest distance of simulated system $r_{0}$=1.12.\\
\indent After the simulation, we know the coordinates of every particle at any time from MD simulated results, but how can we identify the Bravais lattice of simulated system in terms of these coordinates? For one Bravais lattice, the angles between one lattice point and any other two points of its nearest neighbors are constant, so the calculated angles can be used to identify the Bravais lattice. If the distances between one lattice point $i$ and any other two lattice points of its nearest neighbors $i$+1 and $i$+2 are $r_{i,i+1}$ and $r_{i,i+2}$, and the distance between lattice points $i$+1 and $i$+2 is $r_{i+1,i+2}$, then the angle between $\vec{r}_{i,i+1 }$ and $\vec{r}_{i,i+2}$ $\theta$ can be calculated as:
\begin{eqnarray}
\theta=arccos\left (\frac{r^{2}_{i,i+1}+r^{2}_{i,i+2}-r^{2}_{i+1,i+2}}{2r_{i,i+1}r_{i,i+2}} \right)
\end{eqnarray}                  
\indent For ideal fcc lattice, the angles are 60$^{\circ}$, 90$^{\circ}$, 120$^{\circ}$, and 180$^{\circ}$. For ideal hcp lattice ($c/a$$\approx$1.633, $a$ and $c$ being lattice constants), the angles are 60$^{\circ}$, 90$^{\circ}$, 109.5$^{\circ}$, 120$^{\circ}$, 146.4$^{\circ}$, and 180$^{\circ}$. In hcp lattice, the atomic arrangement is described as ABAB$\cdots$, and A and B are the close packed atomic layers. In the A layer or B layer, the angles are 60$^{\circ}$, 120$^{\circ}$, and 180$^{\circ}$. The angles formed between two lattice points in the A layer and one lattice point in the B layer are 60$^{\circ}$ and 90$^{\circ}$, and the angles between one lattice point in the B layer, one lattice point in the A layer on top of the B layer, and one lattice point in the A layer beneath the B layer, are 109.5$^{\circ}$ and 146.4$^{\circ}$. For the particle system, we must calculate all the angles between each lattice point and its nearest neighbors, and then the distribution function $\rho(\theta)$ of these angles between $\theta$ and $\theta$+d$\theta$. For ideal fcc lattice, the distribution function is not zero at the angles of 60$^{\circ}$, 90$^{\circ}$, 120$^{\circ}$, and 180$^{\circ}$, and are all zero at any other angles. The distribution function consists of several discrete points. For ideal hcp lattice, the distribution function is not zero at the angles of 60$^{\circ}$, 90$^{\circ}$, 109.5$^{\circ}$, 120$^{\circ}$, 146.4$^{\circ}$, and 180$^{\circ}$. \\
\indent For the Bravais lattice, the ratios of the distances between one lattice point and any other lattice points are fixed. We can calculate the distances between one lattice point and any other lattice points and compare the calculated results with that of ideal Bravais lattice. With this, we can roughly identify the Bravais lattice. If $r_{i,j}$ is the distance between lattice points $i$ and $j$, we obtain the distances between lattice points $d$
\begin{equation}
d=r_{i,j},j\not=i
\end{equation}                         
\indent After calculating all the values for $d$ for each lattice point, we further calculate the distribution function of the distances between $d$ and $d$+d$d$. With these two distribution functions, we can roughly identify the Bravais lattice of simulated system, but we must further check the particle arrangement for the final identification.
\section{Results and Discussions}
 \begin{figure}[h t b p]
\centering
 \includegraphics[width=80mm]{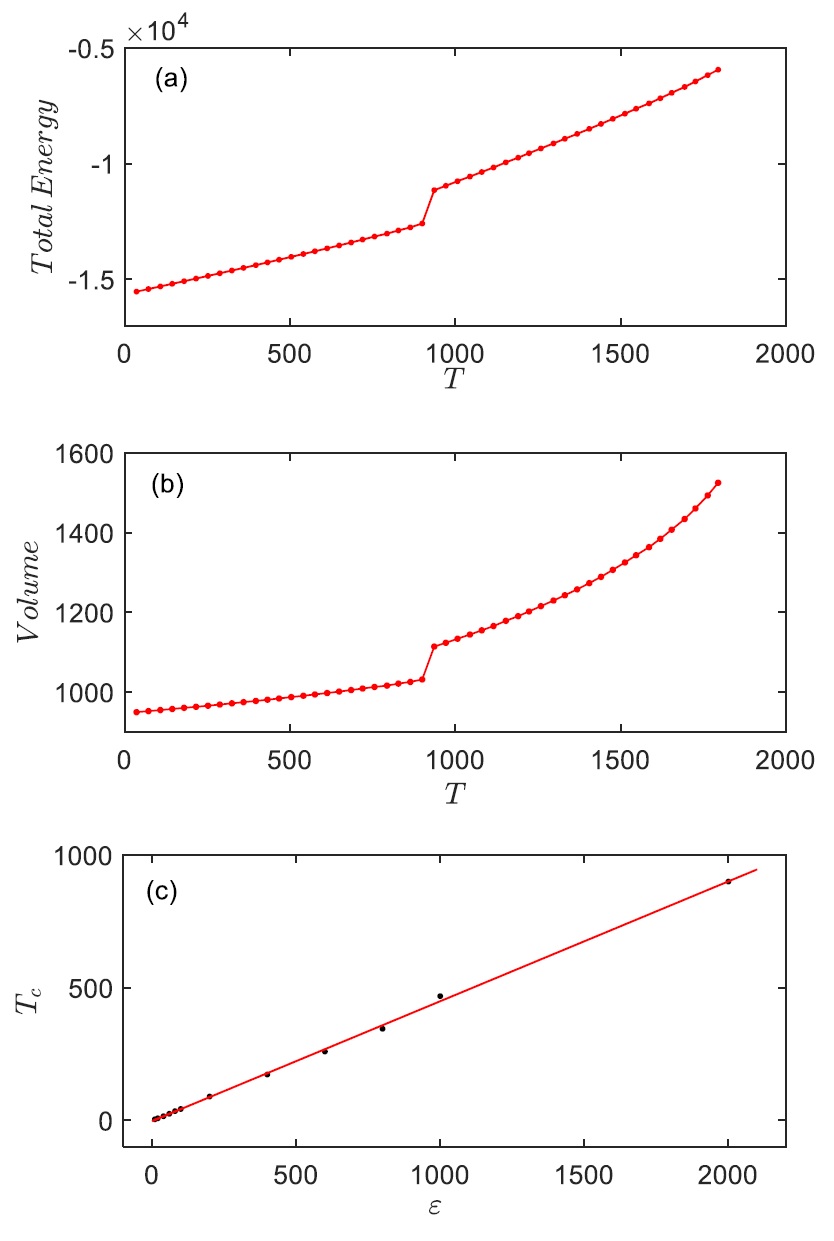}%
 \caption{\label{FIG.1.}For $\epsilon$=2000, dependence of the total energy (a) and volume (b) of simulated system on the temperature, and (c) dependence of the crystallization temperature $T_{c}$ on $\epsilon$.}
 \end{figure}
 \subsection{The phase transition and lattice identification}
We have simulated the crystallization of LJ particles for different $\epsilon$ values. In the simulation, the crystallization temperature $T_{c}$ is defined as the temperature at which the liquid-solid phase transition happens. The equilibrium nearest distance $r_{0}$ is constant for the fixed value of $\sigma$, but the total energy and the crystallization temperature $T_{c}$ of simulated system change for different $\epsilon$ values. Figures 1(a) and 1(b) show the dependence of the total energy and volume of simulated system on the temperature for $\epsilon$=2000. It has been indicated that the total energy and volume decrease as the temperature decreases, and there is an abnormal jump showing the liquid-crystalline phase transition at $T_{c}$. The total energy is in a linear relation with the temperature at the liquid and crystalline states, but the volume is in a nonlinear relation with the temperature only at the liquid state. Figure 1(c) shows the dependence of the crystallization temperature $T_{c}$ on $\epsilon$. As shown in Fig. 1(c), $T_{c}$ is linear to $\epsilon$, indicating that the crystallization temperatures are adjustable. For the particle system with other $\epsilon$ values, the dependence of both the total energy and volume on the temperature is similar to those shown in Figs. 1(a) and (b). Figure 2 shows the particle arrangements of simulated system at the liquid and crystallization states. It has been seen that the atomic arrangement is disordered at the liquid state and ordered at the crystalline state. However, except that, we cannot obtain the information about its crystal lattice and symmetry. \\
 \begin{figure}[h t b p]
\centering
 \includegraphics[width=60mm]{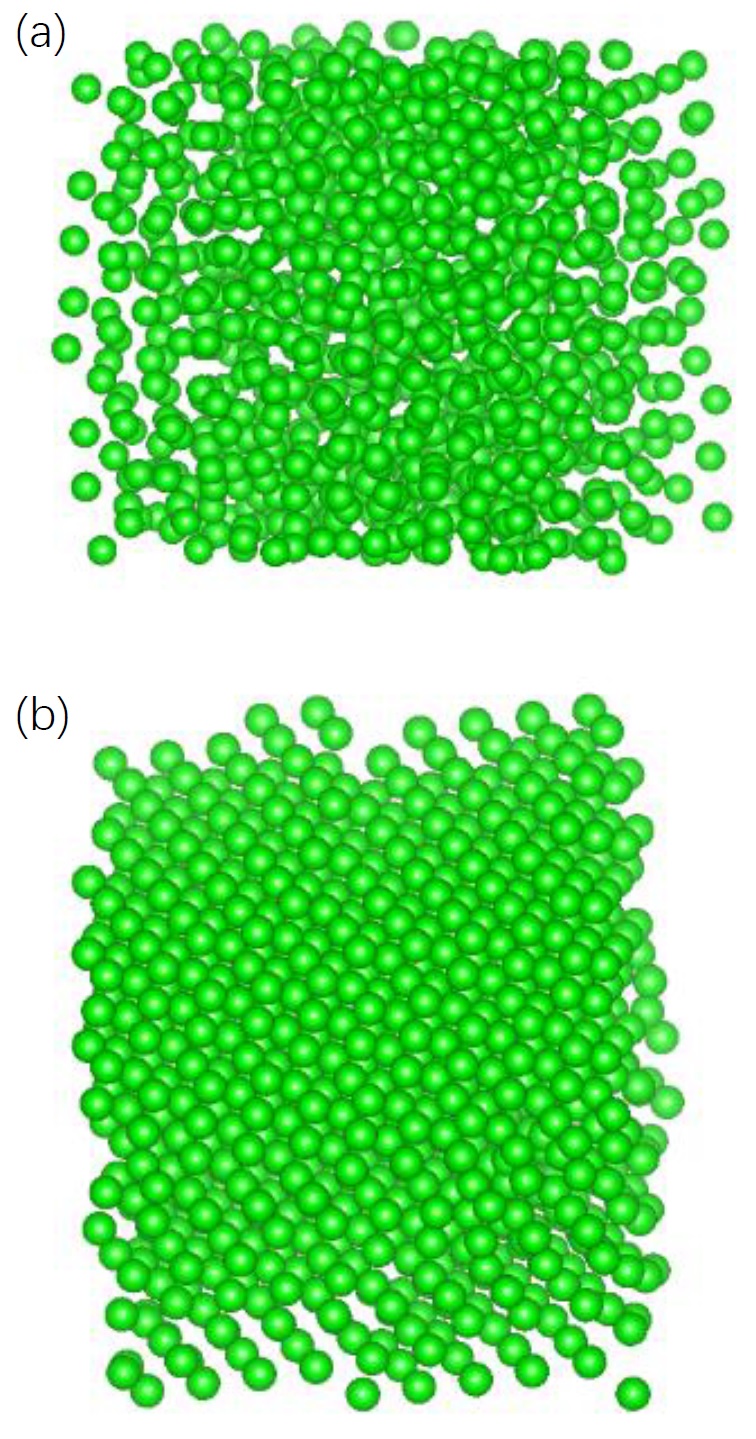}%
 \caption{\label{FIG.2.}For $\epsilon$=2000, the particle arrangements of simulated system at the liquid state (a) and at the crystalline state (b).}
 \end{figure}
 \begin{figure}[h t b p]
\centering
 \includegraphics[width=80mm]{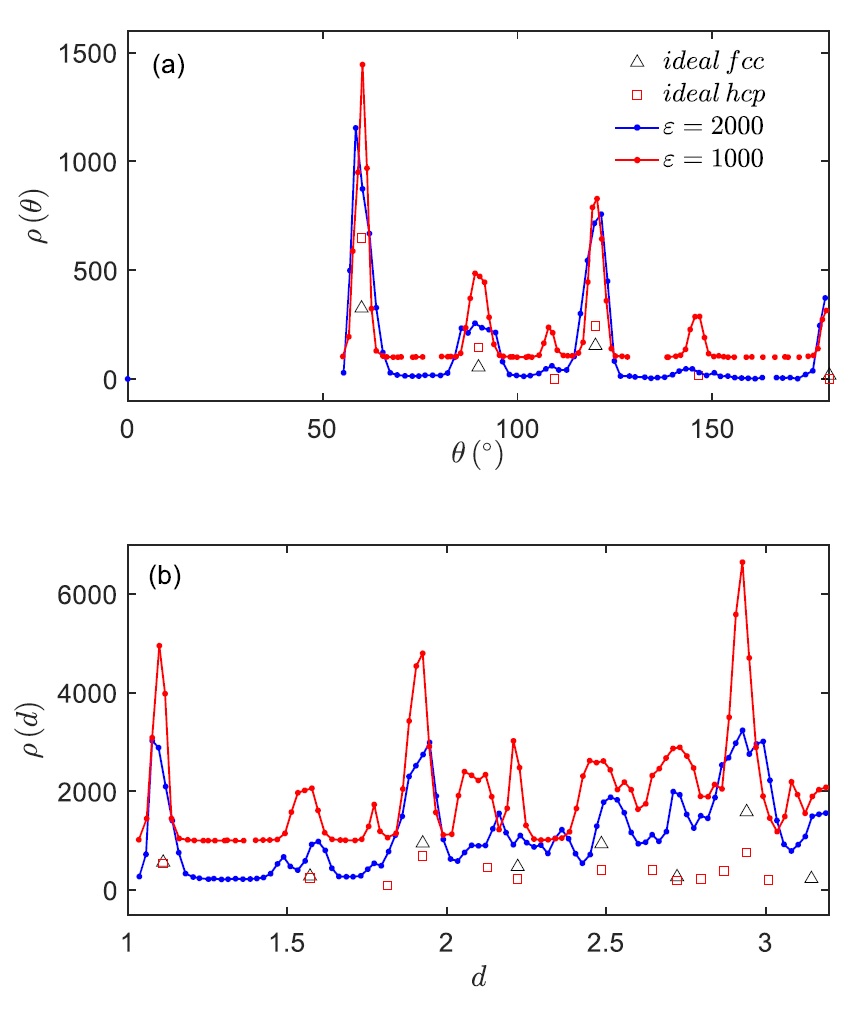}%
 \caption{\label{FIG.3.}For $\epsilon$=1000 and 2000, and $T$=36, the distribution functions of both the angles between one particle and its nearest neighbors (a) and the distances between the particles (b). ‘$\triangle$’ and ‘$\Box$’ denote the distribution functions for both ideal fcc and hcp lattices.}
 \end{figure}
\begin{figure}[h t b p]
\centering
 \includegraphics[width=60mm]{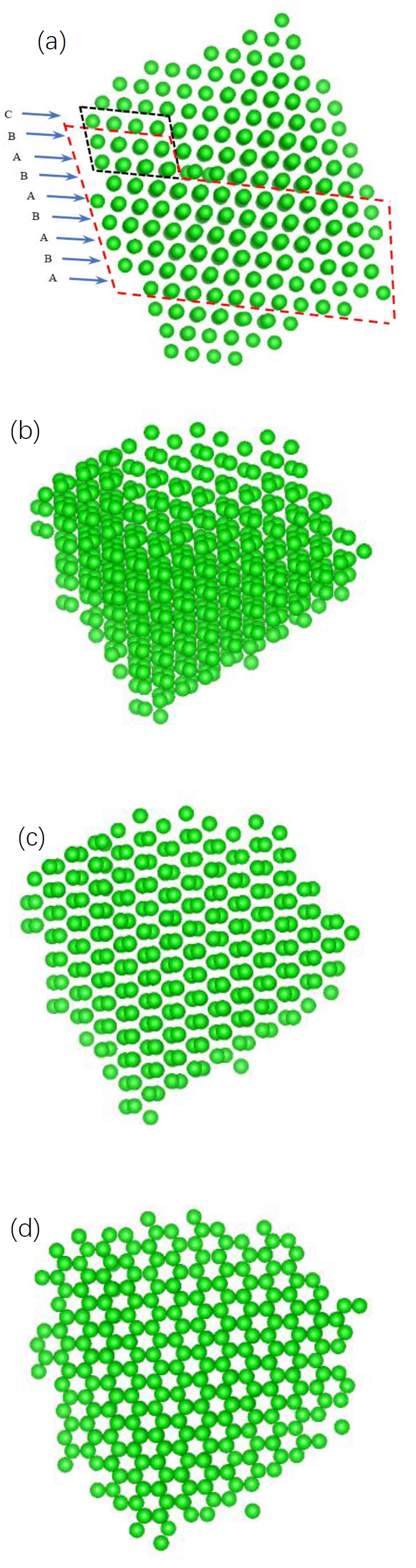}%
 \caption{\label{FIG.4.}For $\epsilon$=1000 and $T$=36, (a) the atomic arrangement of simulated system, (b)-(d) the atomic arrangements of the region encircled by the red dashed lines shown in (a) observed from different directions.}
 \end{figure}
   \begin{figure}[h t b p]
\centering
 \includegraphics[width=60mm]{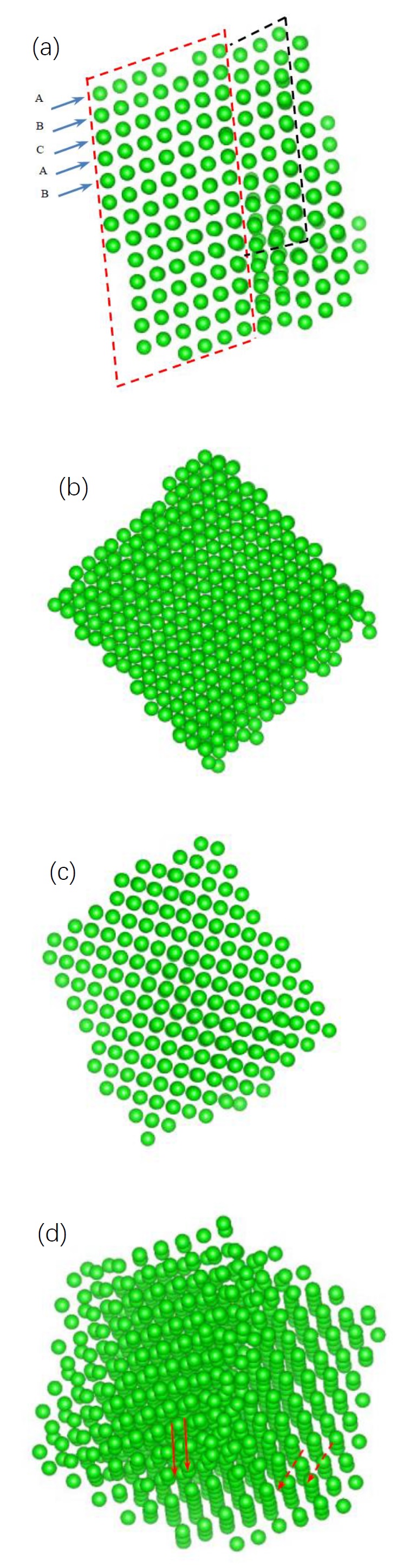}%
 \caption{\label{FIG.5.}For $\epsilon$=2000 and $T$=36, (a) and (d) the atomic arrangements of simulated system observed from different directions, and (b) and (c) the atomic arrangements of the region encircled by the red dashed lines shown in (a) observed from the directions normal to (111) and (100) planes.}
 \end{figure}
\indent Figure 3 shows the distribution functions of both the angles between one particle and its nearest neighbors and the distances between the particles for $\epsilon$=1000 and 2000, and $T$=36. It has been shown from Fig. 3(a) that for $\epsilon$=1000 and 2000 at the crystalline state $\rho$($\theta$) is in good agreement with that of ideal hcp lattice. However, the values corresponding to the angles of 109.5$^{\circ}$ and 146.4$^{\circ}$ for $\epsilon$=1000 are much larger than those for $\epsilon$=2000. It has also been seen from Fig. 3(b) that ideal fcc lattice is very similar to ideal hcp lattice because their main peaks correspond to the same distances. But due to the significant difference between the atomic arrangements, ideal hcp lattice has more weak peaks than ideal fcc lattice does. Our simulated results are in good agreement with those of both ideal lattices for the first several main peaks so we must check the atomic arrangement of simulated system for the final identification. Figure 4 shows the atomic arrangements of simulate system. It has been seen from Fig. 4(a) that our simulated system shows the well-defined atomic arrangement. The atomic arrangement in a region encircled by the red dashed lines is ABABABAB, and the same as that in hcp lattice. There is also a region encircled by the black dashed lines showing the ABC atomic arrangement which is the same as that in fcc lattice. The region on top of the red dashed lines shows the ABAB$\cdots$ atomic arrangement again, and the same as that in the region encircled by the red dashed lines. However, the transition region between these two regions shows the wrong atomic arrangement of ABC. Next, we remove all the particles outside the red dashed lines and check whether the lattice is the hcp lattice or not from different observed directions. It has been shown in Fig. 4(b) that the lattice shows clearly the ABAB$\cdots$AB atomic arrangement. As shown in Fig. 4(c), due to the small displacement of A layer particles with respect to B layer particles, there is a partial overlap of A layer on B layer when observed from the direction normal to A layer. The atomic arrangement in Fig. 4(d) shows the typical characteristic of hcp lattice, and its crystal symmetry. Till now, we can identify our simulated system for $\epsilon$=1000 showing hcp lattice.\\
\indent We must also check the atomic arrangement of simulated system for the final identification of its lattice for $\epsilon$= 2000. Figure 5 shows the atomic arrangements of our simulated system at the crystalline state. The atomic arrangement of simulated system observed from some direction is shown in Fig. 5(a). It has been seen that there is a small region in which the wrong atomic arrangement occurs on the right bottom of the system, and the expected full overlap is not good. Meanwhile, there is a small difference between the atomic arrangements of different regions, for instance, that between regions encircled by the red dashed lines and black dashed lines. The region encircled by the red dashed lines shows the atomic arrangement of ABCABC$\cdots$ABC, which is the same as that of fcc lattice. The atomic arrangements of simulated system observed from the directions normal to (111) and (100) planes are shown in Figs. 5(b) and 5(c), respectively, and the same as that of ideal fcc. Now, we can identify the system showing fcc lattice for $\epsilon$=100. The deviation from ideal fcc lattice shown in Fig. 3 stems from the wrong atomic arrangements of the system. As shown in Fig. 5(d), two regions showing the fcc lattice have a small relative displacement with respect to each other, as indicated by red solid arrows. However, the displacement is the same as that of A layer with respect to B layer, and this leads to the AB atomic arrangement at the transition region. In the meantime, there is a small fraction of the unwanted ABAB atomic arrangements in the system showing fcc lattice, as shown by red dashed arrows in Fig. 5(d). These mismatches lead to the difference between our system and ideal fcc lattice. From the results above, we can roughly identify the Bravais lattice by using $\rho$($\theta$) and $\rho$($d$).\\
\begin{figure}[h t b p]
\centering
 \includegraphics[width=80mm]{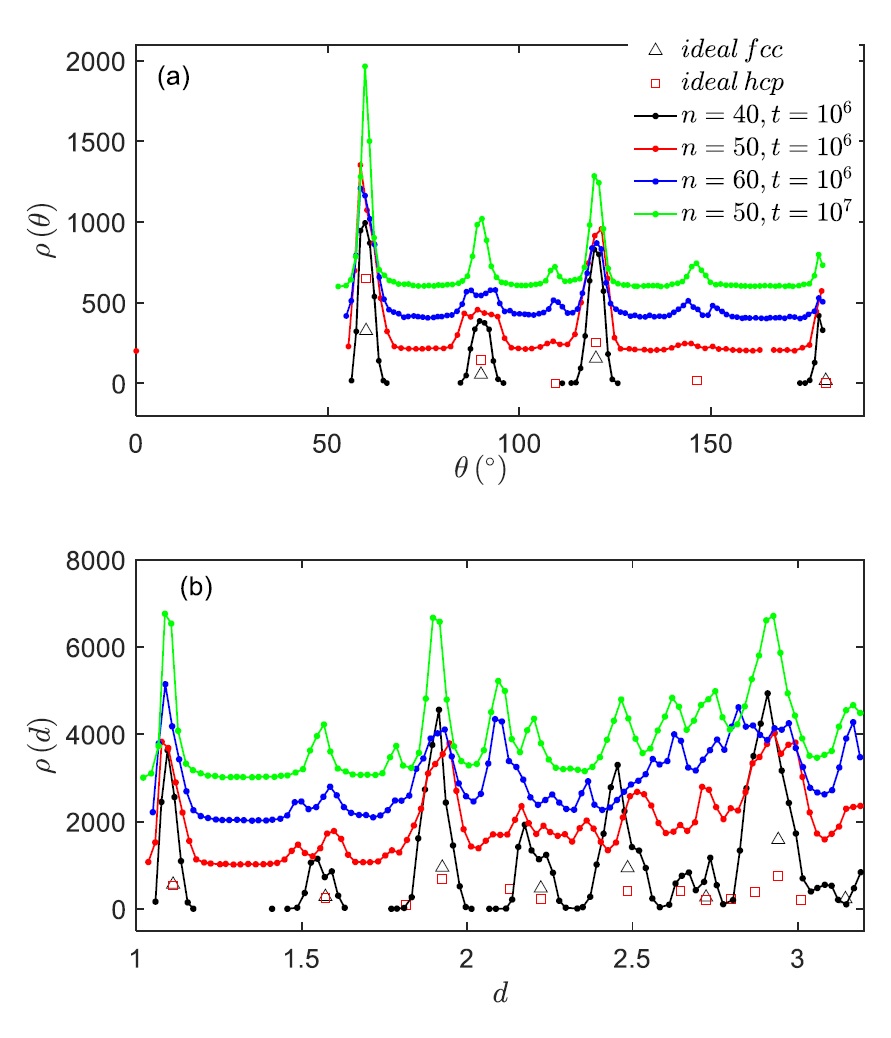}%
 \caption{\label{FIG.6.}For $\epsilon$=2000 and $T$=36, the distribution functions of both the angles between one particle and its nearest neighbors (a) and the distances between the particles (b) with different annealing temperature step $n$ and annealing time $t$. ‘$\triangle$’ and ‘$\Box$’ denote the distribution functions for both ideal fcc and hcp lattices.}
 \end{figure}
\begin{figure*}%[h t b p]
\centering
 \includegraphics[width=160mm]{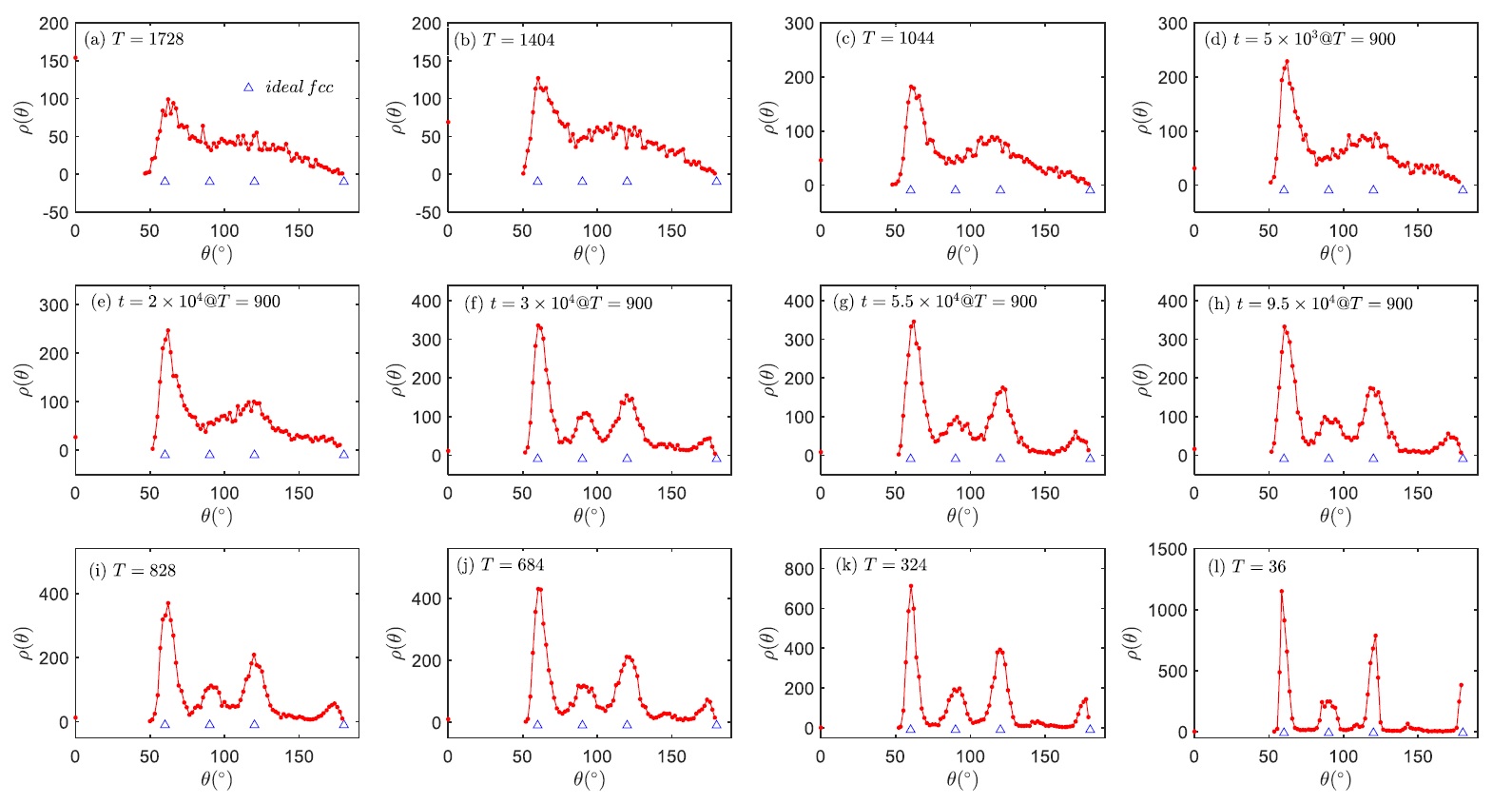}%
 \caption{\label{FIG.7.}For $\epsilon$=2000, the distribution functions of the angles between one particle and its nearest neighbors at different temperatures and different annealing time. ‘$\triangle$’ denotes the distribution functions for ideal fcc lattice.}
 \end{figure*} 
 \begin{figure*}%[h t b p]
\centering
 \includegraphics[width=160mm]{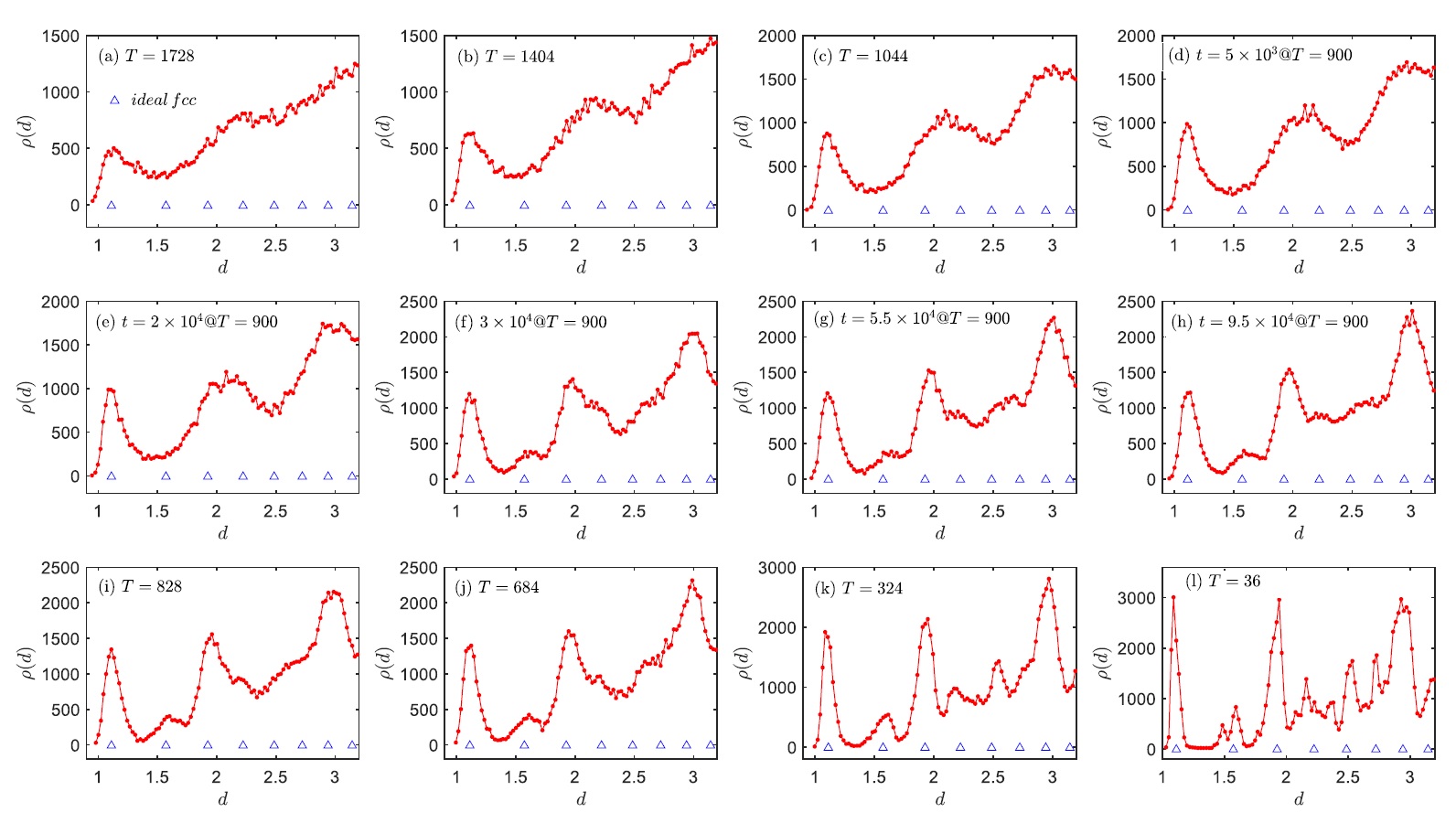}%
 \caption{\label{FIG.8.}For $\epsilon$=2000, the distribution functions of the distances between the particles at different temperatures and different annealing time. ‘$\triangle$’ denotes the distribution functions for ideal fcc lattice.}
 \end{figure*} 
\indent We have investigated the systems with $\epsilon$=10, 20, 40, 60, 80, 100, 200, 400, 600, 800, 1000, and 2000. We expect that for every $\epsilon$ value, simulated system should show the only Bravais lattice which will be insensitive to simulation parameters. The results have shown that this is not the case. The systems show hcp lattice for $\epsilon$=40, 80, 200, 800, and 1000, and fcc lattice for $\epsilon$=10, 20, 60, 100, 400, 600, and 2000. These fcc systems show the significant wrong atomic arrangements and this is in agreement with the results in Ref. [5]. We have also investigated the influence of the annealing temperature step $n$ and the annealing time $t$ on the lattice of the system for $\epsilon$=2000. Figure 6 shows the distribution functions of both the angles between one particle and its nearest neighbors and the distances between the particles with different annealing temperature step $n$ and annealing time $t$ for $\epsilon$=2000 and $T$=36. From Fig. 6(a) and the atomic arrangements of the systems, it has been found that for annealing time $t$=1$\times$10$^{6}$, the systems for $n$=40 and 50 show fcc lattice, and the one for $n$=60 shows hcp lattice. For $n$=50, the system with $t$=1$\times$10$^{6}$ shows fcc lattice, and the one with $t$=1$\times$10$^{7}$ shows hcp lattice. This means that the lattice of simulated system is greatly sensitive to dynamic parameters, and the system may show fcc lattice or hcp lattice. Our simulated results have confirmed earlier investigations \cite{Kihara-2,Barron-3}. The small energy difference between the hcp and fcc lattices has an effect on their existence but the hcp lattice is more stable than the fcc lattice. However, in the simulation, we cannot predict which lattice the system will show. This also means that the application of LJ potential is limited.\\
\begin{figure}[h t b p]
\centering
 \includegraphics[width=60mm]{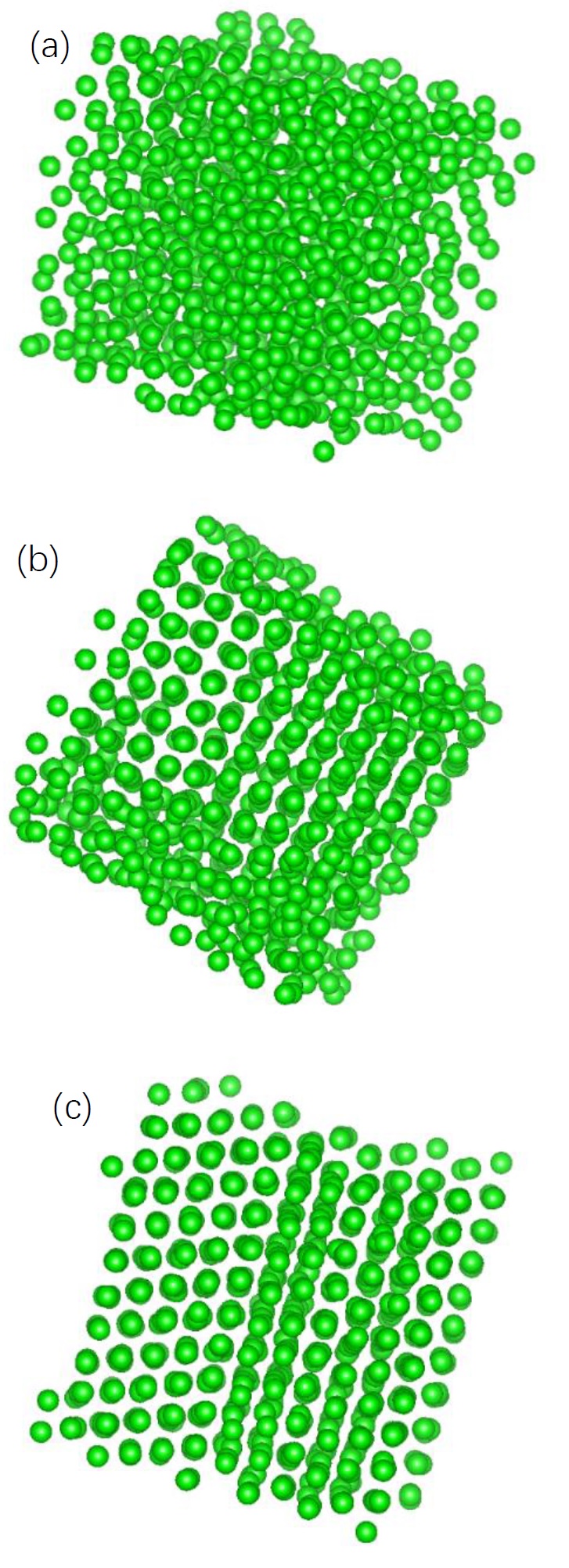}%
 \caption{\label{FIG.9.}The atomic arrangements of simulated system at the beginning of the crystallization for $\epsilon$=2000 and $T$=900. (a)2$\times$10$^{4}$@$T$=900, (b)2.5$\times$10$^{4}$@$T$=900, and (c)3$\times$10$^{4}$@$T$=900.}
 \end{figure}
 \begin{figure}[h t b p]
\centering
 \includegraphics[width=80mm]{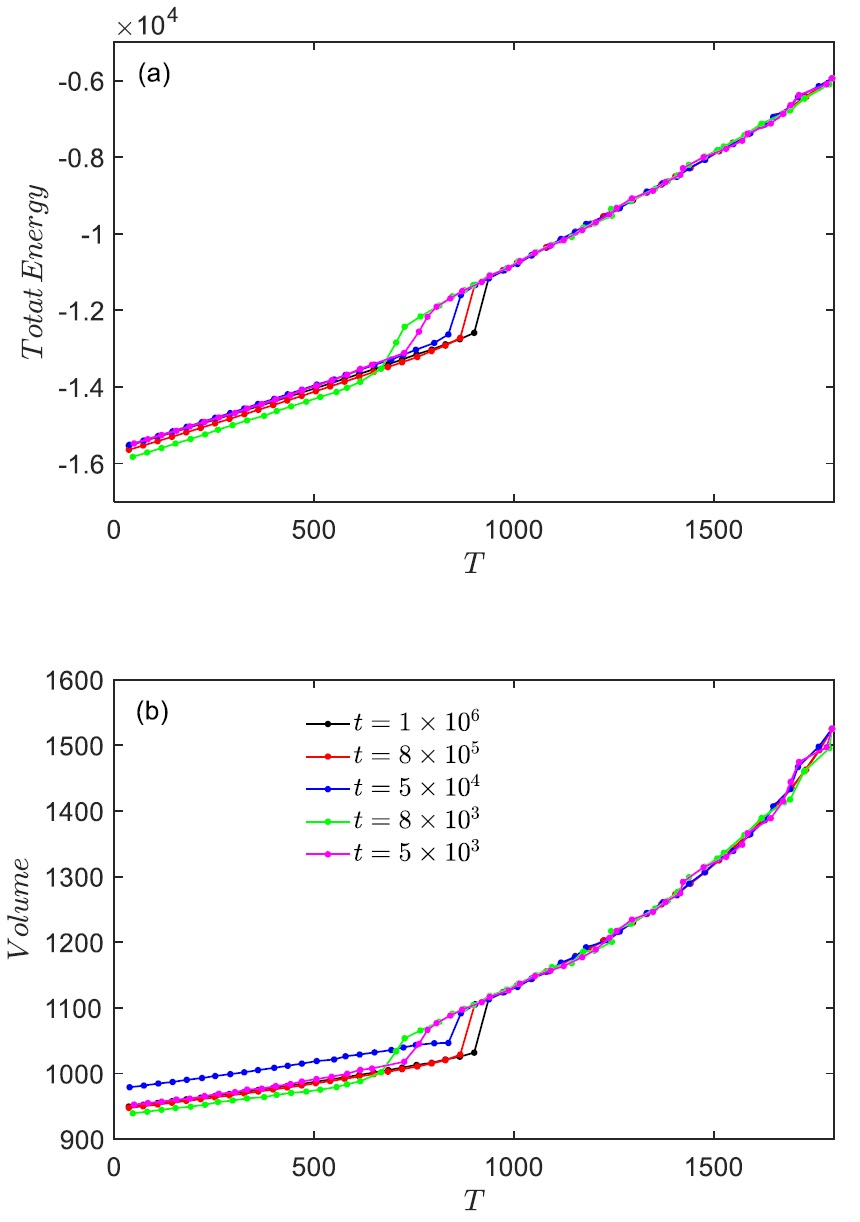}%
 \caption{\label{FIG.10.}Dependence of the total energy (a) and volume (b) of simulated system on the temperature for $\epsilon$=2000 with different annealing time $t$.}
 \end{figure}
  \begin{figure}[h t b p]
\centering
 \includegraphics[width=80mm]{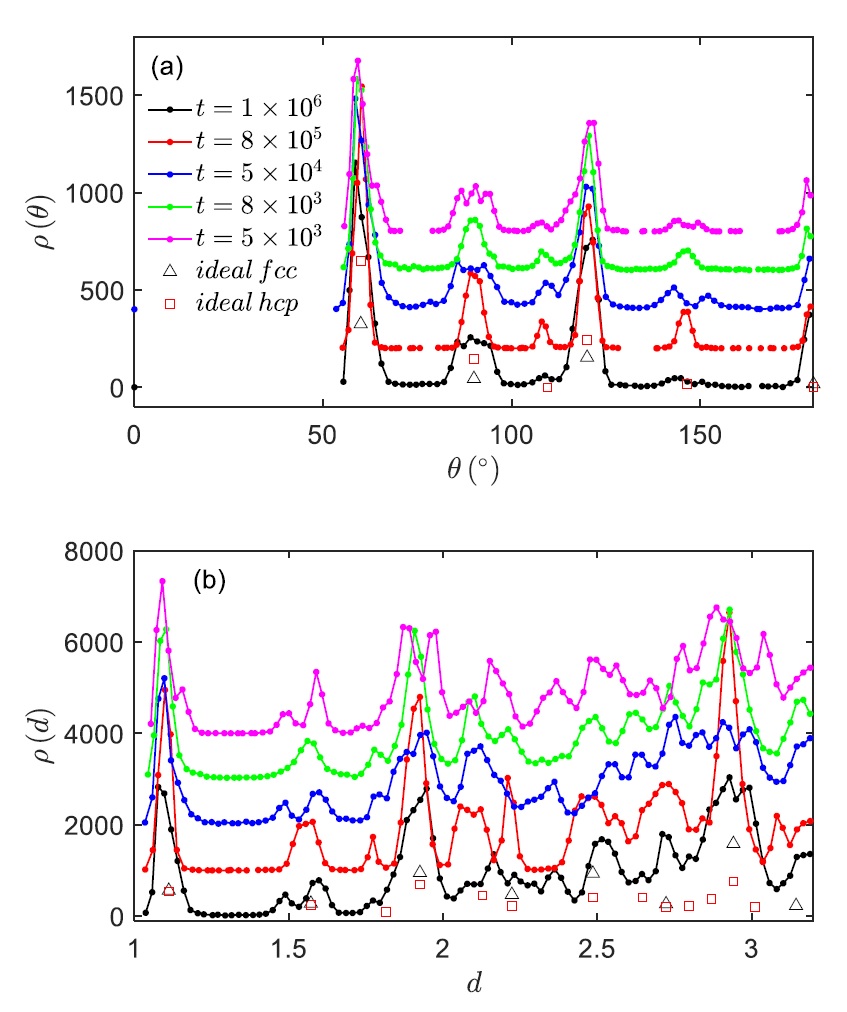}%
 \caption{\label{FIG.11.}The distribution functions of both the angles between one particle and its nearest neighbors (a) and  the distances between the particles (b) with different annealing time $t$ for $\epsilon$=2000. ‘$\triangle$’ and ‘$\Box$’ denote the distribution functions for both ideal fcc and hcp lattices.}
 \end{figure}
 \subsection{Evolution of microstructure of simulated system}
\indent According to the calculated results for $\rho$($\theta$) and $\rho$($d$), we can investigate both the microstructure of the liquid and the abnormal transition of the particle arrangements at crystallization temperature $T_{c}$. Figures 7 and 8 show the distribution functions of both the angles between one particle and its nearest neighbors, and the distances between the particles at different temperatures and different annealing time for $\epsilon$=2000. From Figs. 7(a)-(c), it has been demonstrated that at the liquid state $\rho$($\theta$) show a peak at the angle of 60$^{\circ}$, and $\rho$($\theta$) are zero for $\theta$$<$50$^{\circ}$,and not zero for $\theta$$>$50$^{\circ}$. With decrease of the temperature, the value for $\rho$($\theta$) at 60$^{\circ}$ also increases, and $\rho$($\theta$) show another weak peak at the angle of 120$^{\circ}$. As shown in Figs. 8(a)-(c), $\rho$($d$) are zero for $d$$<$1, and this means that the distances between particles cannot be too close and far away from its equilibrium nearest distance $r_{0}$=1.12, or this could cause a significant increase in the total energy. In contrast with the distribution functions of ideal fcc lattice, there is a clear peak at the first distance $d_{1}$. With decreasing the temperature, there are two broad peaks corresponding to the third and fourth distances ($d_{3}$ and $d_{4}$) and the sixth, seventh, and eighth distances ($d_{6}$, $d_{7}$, and $d_{8}$), respectively, but there are no clear peaks occurring at the second distance ($d_{2}$) and the fifth distance ($d_{5}$). In the simulation, we carried out the NPT operations for the annealing time of 1$\times$10$^{6}$ for each temperature. At the liquid state, the distribution functions $\rho$($\theta$) and $\rho$($d$) at each temperature do not change significantly. The distribution functions $\rho$($\theta$) and $\rho$($d$) of simulated system at the crystallization temperature $T_{c}$ are demonstrated in Figs. 7(d)-(h) and Figs. 8(d)-(h). At the initial stage when the temperature decreases to the crystallization temperature $T_{c}$=900, for example, at $t$=5$\times$10$^{3}$@$T$=900 and 2$\times$10$^{4}$@$T$=900, $\rho$($\theta$) and $\rho$($d$) of simulated system do not show any significant change when contrasting with these of the liquid. However, at $t$=3$\times$10$^{4}$ @$T$=900, $\rho$($\theta$) and $\rho$($d$) both show clear changes. It has been seen in Fig. 7(f) that there are two clear peaks at the angles of 90$^{\circ}$ and $\sim$180$^{\circ}$. That is, all the peaks of $\rho$($\theta$) which are required by ideal fcc lattice are present. As shown in Fig. 8(f), there is also a clear peak corresponding to the second distance ($d_{2}$). As shown in Figs. 7(e)-(h), $\rho$($\theta$) do not show any significant change. In Figs. 8(e)-(h), the value for $\rho$($d$) corresponding to the second distance ($d_{2}$) also do not change significantly, but there are two clear peaks corresponding to the third distance ($d_{3}$) and the seventh distance ($d_{7}$). In Figs. 7(i)-(l) and Figs. 8(i)-(l), the values for $\rho$($\theta$) and $\rho$($d$) increase and the corresponding peaks become sharper as the temperature further decreases. For the temperatures below $T_{c}$, the values for $\rho$($\theta$) and $\rho$($d$) do not change significantly as the annealing time increases. From the results above, it has been indicated that whether there are clear sharp peaks corresponding to the angle of 90$^{\circ}$ and the second distance ( $d_{1}$) or not is very important for the crystallization. The second distance is defined as the distance between one particle and its second nearest neighbors. If the particle, its nearest neighbors, and its second nearest neighbors are arranged to form an fcc lattice, then there are clear peaks corresponding to the angle of 90$^{\circ}$ and the second distance ($d_{2}$). This means that the fcc lattice has formed, and it must be reflected in the atomic arrangement of simulated system. Figure 9 shows the atomic arrangements of simulated system at the beginning of the crystallization for $\epsilon$=2000 and $T$=36. Fig. 9(a) shows the atomic arrangement corresponding to Figs. 7(e) and 8(e). It has been seen that there are no peaks corresponding to the angle of 90$^{\circ}$ and the second distance ($d_{2}$), and the system is disordered. As shown in Fig. 9(c), the system is ordered. This corresponds to the cases shown in Figs. 7(f) and 8(f) and there are clear peaks. Fig. 9(b) shows the atomic arrangement corresponding to the intermediate stage between the cases described in Figs. 7(e) and 8(e). It has been found that a fraction of the system is ordered and the remaining fraction is still disordered. This means that the formation of the order of the second nearest neighbors plays a key role in the crystallization. This disorder-order phase transition of the atomic arrangements can be rapidly completed in a very short time at the crystallization temperature.\\
 \begin{figure}[h t b p]
\centering
 \includegraphics[width=80mm]{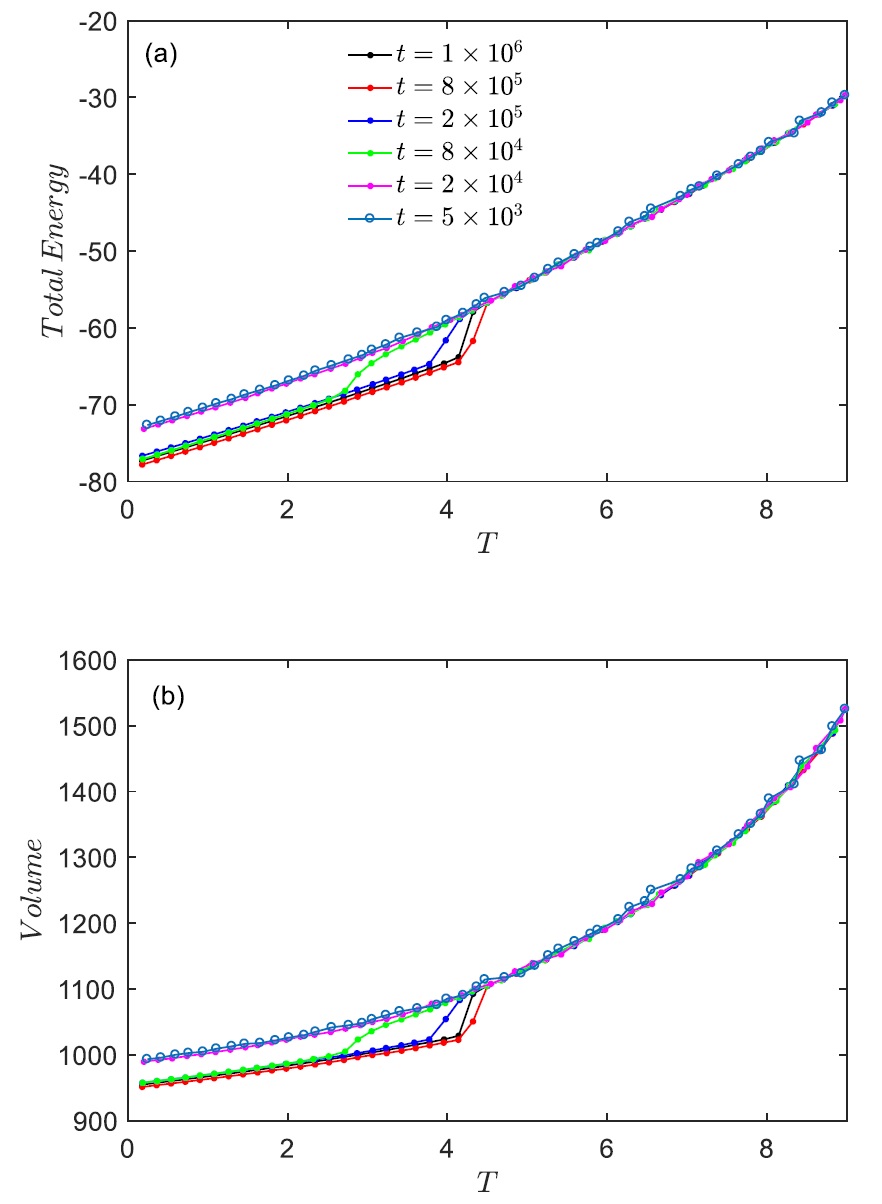}%
 \caption{\label{FIG.12.}Dependence of the total energy (a) and volume (b) of simulated system on the temperature for $\epsilon$=10 with different annealing time $t$.}
 \end{figure}
  \begin{figure}[h t b p]
\centering
 \includegraphics[width=80mm]{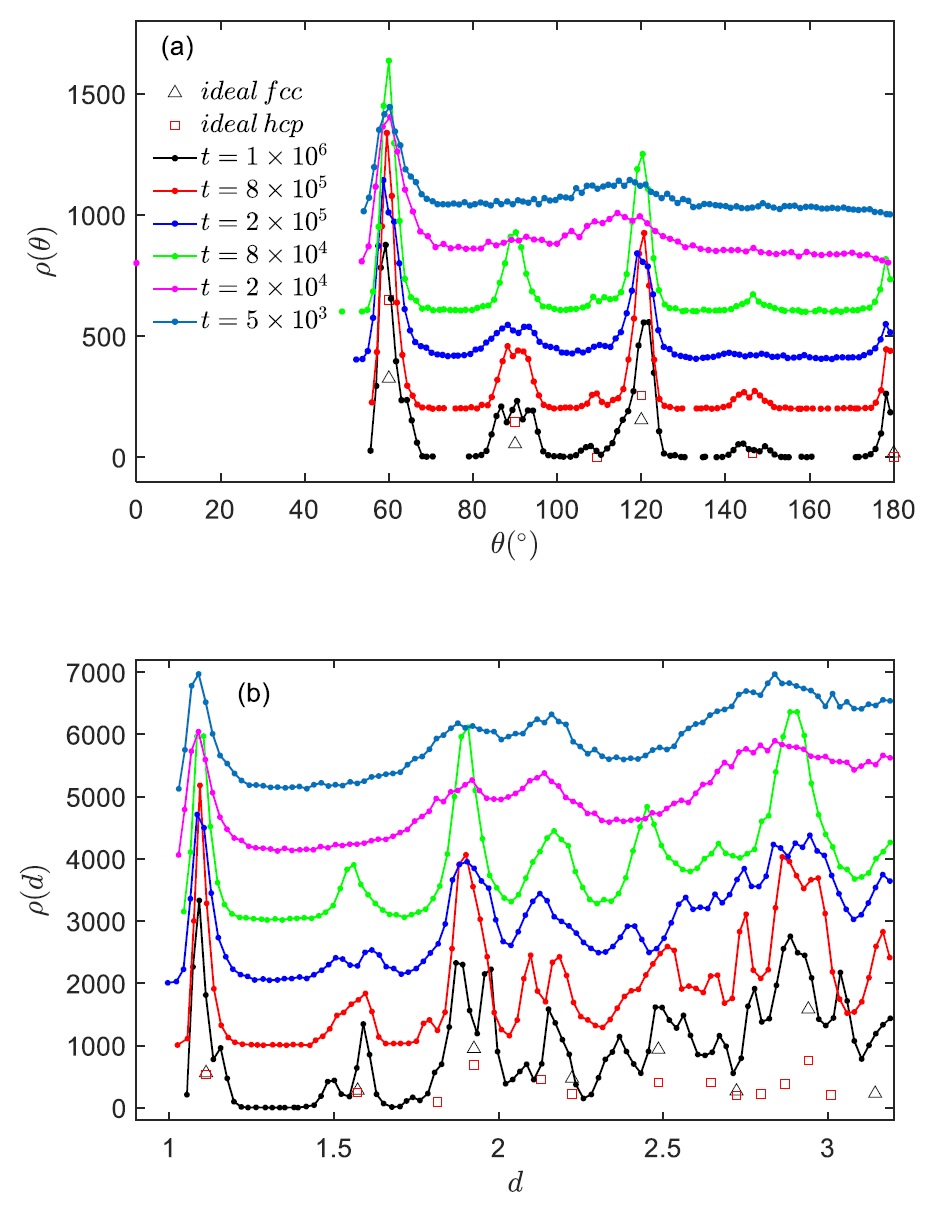}%
 \caption{\label{FIG.13.}The distribution functions of both the angles between one particle and its nearest neighbors (a) and  the distances between the particles (b) with different annealing time $t$ for $\epsilon$=10. ‘$\triangle$’ and ‘$\Box$’ denote the distribution functions for both ideal fcc and hcp lattices.}
 \end{figure}
\subsection{The formation of Non-crystalline system}
\indent With LJ potential, we expect that the liquid state of the system can be preserved by decreasing the annealing time. We chose the systems with $\epsilon$=2000 and $\epsilon$=10. Figure 10 shows the dependence of the total energy and volume of simulated system on the temperature for $\epsilon$=2000 with different annealing time $t$. From Fig. 10, it has been seen that the dependence of the total energy and the volume of the system on the temperature also change clearly and the crystallization temperature decreases gradually as the annealing time decreases. However, the system still shows the crystallization state even if the annealing time is reduced to 5$\times$10$^{3}$. Figure 11 shows the distribution functions of both the angles between one particle and its nearest neighbors and the distances between the particles with different annealing time $t$ for $\epsilon$=2000. From Fig. 11, it has been found that the distribution functions $\rho$($\theta$) and $\rho$($d$) of the system are in good agreement with that of ideal hcp lattice. For example, at $t$=8$\times$10$^{5}$ and 8$\times$10$^{4}$, there are clear peaks at the angles of 109.5$^{\circ}$ and 146.4$^{\circ}$. We checked the atomic arrangements and found that those systems still show the fcc lattice and there are heavy wrong atomic arrangements.\\
\indent Figure 12 shows the dependence of the total energy and volume of simulated system on the temperature for $\epsilon$=10 with different annealing time $t$. As shown in Fig. 12, the results are similar to those for $\epsilon$=2000 shown in Fig. 10, but we successfully preserved the liquid state of the system. At $t$=2$\times$10$^{4}$, there is no abnormal jump but the gradual change of the total energy (or volume) of the system as the temperature decreases. This is in good agreement with the data reported in the literature \cite{Debenedetti-9,Shintani-10}. Figure 13 shows the distribution functions of both the angles between one particle and its nearest neighbors and the distances between the particles with different annealing time $t$ for $\epsilon$=10. Combining with those shown in Fig. 13, we also checked the atomic arrangements of the systems, and found that the systems show the fcc lattice for $t$=8$\times$10$^{5}$, 2$\times$10$^{5}$, and 8$\times$10$^{4}$. For $t$=2$\times$10$^{4}$ and 5$\times$10$^{3}$, there are no corresponding peaks such as the peaks at the angle of 90$^{\circ}$ and at the second distance ($d_{2}$). The microstructures of the systems are similar to that of the liquid shown in Fig. 7(e). The non-crystalline system shows no order of the second nearest neighbors, in comparison with the crystalline state. From the results above, it implies that large values for $\epsilon$ are helpful for the crystallization and small values for $\epsilon$ for the non-crystallization.\\
\indent Our results have shown that without setting any initial Bravais lattice such as fcc and hcp lattices, with the simple LJ potential, and by randomly creating the particles in the simulation box, we successfully simulated the crystallization and non-crystallization of LJ particles. We found the systems showing fcc lattice or hcp lattice, and we also preserved the non-crystalline state. These results are very interesting because for many decades the underlying physics about the crystallization is always thought to be very complex. The fact that the complicated fcc lattice and hcp lattice can be formed by MD simulation with the simple LJ potential, implies that the interatomic potential may be very simple and is not as complicated as we might think before. In our simulation, with LJ potential, we cannot expect a fixed and predicable Bravais lattice. However, LJ potential can be a starting point from which we can find a new interatomic potential for a fixed and predicable lattice such as fcc, hcp, and body-centered cubic (bcc) lattices. Such a research is now under way.\\
\indent We have also introduced new distribution functions $\rho$($\theta$) and $\rho$($d$) to roughly identify Bravais lattice of the system. But we must further check the atomic arrangements for the final identification of the system. For example, as shown in Fig. 13, the system shows the hcp lattice from the distribution functions $\rho$($\theta$) and $\rho$($d$). However, the system shows the fcc lattice just because of the heavy wrong atomic arrangements. In all, these new distribution functions are important for our simulation, and they have played a role like x-ray diffraction (XRD).\\
\indent It must be pointed out that in this paper, we limit our focus on the realization of the simulation and the characterization of simulated system, and pay little attention to other related physical issues. However, our simulation is helpful for investigating many fundamental issues in condensed physics and material science. For example, the atomic arrangements shown in Figs. 7 and 8 are useful for studying the growing mechanism of the crystal, and also helpful to investigate the defects in crystal such as grain boundaries. Therefore, our simulation, without setting any initial Bravais lattice, and with the simple interatomic potential, can greatly extend the application of MD simulation, and allow a deep understanding of the microstructure of material. In the meantime, due to the simplicity of LJ potential, the computation cost can be greatly reduced. Therefore, we can carry out MD simulation of large scale particle system that can consist of more than one million particles, and to some degree, this implies that an era in which we can reveal the microstructure of material and its evolution with MD simulation of large scale particle system is coming.\\
%\label{}
%\subsection{}
%\subsubsection{}
%\section{THEORETICAL MODELING}
%\label{}
%\subsection{}
%\subsubsection{THEORETICAL MODELING}
\section{Conclusions}
Without setting any initial Bravais lattice, and with the simple LJ potential, we successfully reproduced the crystallization and non-crystallization of LJ particles by MD simulation. In the simulation, our simulated systems show fcc lattice and hcp lattice, and we also preserved the liquid state of the system. The microstructure of the non-crystalline is similar to that of the liquid near the crystallization temperature, and there is no order of the second nearest neighbors when compared with the crystallization. The distribution functions of both the angles between one particle and its nearest neighbors, and the distances between the particles can be used to roughly identify the Bravais lattice of the system, and the atomic arrangement must be checked for the final identification of the system. The systems are very sensitive to dynamic parameters and show either fcc lattice or hcp lattice. Our simulation is helpful for investigating the fundamental issues in condensed physics and material science.
%\begin{widetext}
%$$\mbox{put long equation here}$$
%\end{widetext}
% Figures should be put into the text as floats. 
% Use the graphics or graphicx packages (distributed with LaTeX2e).
% Put the label that you will use with \ref{} command in the braces of the \label{} command.
% Insert the column specifiers (l, r, c, d, e$T_{c}$.) in the empty braces of the
% \begin{table}
% \end{tabular}
% \end{table}
\appendix
\section{in script}
\begin{lstlisting}
units         lj
boundary      p p p
atom_style    atomic
dimension     3
region        box block 0 10 0 10 0 10
create_box    1 box
create_atoms  1 random 1000 245 box
timestep      0.001
thermo        1000
group         big1 type 1
mass          1 100
pair_style    lj/cut 2.5
pair_coeff    1 1 20 1.0 2.5
minimize      1.0e-10 1.0e-10 1000000 &
              1000000
run           1000
variable      ltem equal 18
velocity      all create ${ltem} 314029 &
              loop geom
fix           1 all npt temp ${ltem} &
              ${ltem} 1.0 iso 0.0 0.0 1.0
dump          1 all xyz 5000 file1.*.xyz
run           1000000
undump        1
variable      lpa equal ${ltem}
label         loopa
variable      a loop 49
variable      tem equal ${lpa}-0.36*$a
fix           1 all npt temp ${tem} &
              ${tem} 1.0 iso 0.0 0.0 1.0
dump          1 all xyz 5000 file1.*.xyz
run           1000000
undump        1
next          a
jump          SELF loopa
\end{lstlisting}
\begin{acknowledgments}
This work is supported by the National Natural Science Foundation of China (Grant No. 11204087).
% Put your acknowledgments here.
\end{acknowledgments}

\end{document}